\documentclass[a4paper,11pt]{article}
\usepackage{aaskaiid}
\usepackage{orcidlink}
\title{The Nearest Galactic Nucleus: Studying the Galactic Centre with SKA-Mid}
\ShortTitle{The Galactic Centre}

\author[1]{Rainer Sch{\"o}del\orcidlink{0000-0001-5404-797X}}
\ShortName{Sch{\"o}del et al.} % shortened name list for header 
\author[1]{Antxon Alberdi\orcidlink{0000-0002-9371-1033}}
\author[2]{Izaskun Jim\'enez-Serra\orcidlink{0000-0003-4493-8714}}
\author[4]{Michael Kramer\orcidlink{0000-0002-4175-2271}}
\author[3]{Farhad Yusef-Zadeh\orcidlink{0000-0001-8551-9220}}
\author[1]{Miguel P\'erez-Torres\orcidlink{0000-0001-5654-0266}}
\author[5]{Mark R. Morris\orcidlink{0000-0002-6753-2066}}
\author[6]{Rob Fender\orcidlink{0000-0002-5654-2744}}
\author[7]{Jan Forbrich\orcidlink{0000-0001-8694-4966}}
\author[8]{Adriano Ingallinera\orcidlink{0000-0002-3137-473X}}
\author[1]{Miguel Cano Gonz\'alez\orcidlink{0000-0002-9664-7700}}
\author[9]{Jonathan D. Henshaw\orcidlink{0000-0001-9656-7682}}
\author[9,10]{Steven N. Longmore\orcidlink{0000-0001-6353-0170}}
\author[1]{Javier Mold\'on\orcidlink{0000-0002-8079-7608}}
\author[1]{Angela Gardini\orcidlink{0000-0002-3234-9130}}
\author[11,12,13]{Ian Heywood\orcidlink{0000-0001-6864-5057}}
\author[4]{Isabella Rammala-Zitha\orcidlink{0000-0002-0224-6579}}
\author[14]{Fatemeh S.~Tabatabaei\orcidlink{0000-0002-0377-0970}}
\author[14]{Farideh Mazoochi\orcidlink{0000-0003-3390-4893}} 
\author[4,15]{Veena Vadamattom Shaji\orcidlink{0000-0002-0801-8550}}
\author[16]{Alessio Traficante\orcidlink{0000-0003-1665-6402}}
\author[17]{Michal Zaja\v{c}ek\orcidlink{0000-0001-6450-1187}}
\author[18]{Jaroslav Haas\orcidlink{0000-0001-8594-3689}}
\author[1]{Lourdes Verdes-Montenegro\orcidlink{0000-0003-0156-6180}}
\author[1]{Susana S\'anchez Exp\'osito\orcidlink{0000-0002-7510-7633}}
\author[1]{\'Alvaro Mart\'inez Arranz\orcidlink{0000-0002-8524-632X}}
\author[1]{Francisco Nogueras-Lara\orcidlink{0000-0002-6379-7593}}

\affiliation[1]{Instituto de Astrof\'isica de Andaluc\'ia (CSIC),
  Glorieta de la Astronom\'ia s/n, 18008 Granada, Spain}
\emailAdd{rainer@iaa.es}
\affiliation[2]{Centro de Astrobiolog\'ia (CAB), CSIC-INTA, Torrej\'on de Ardoz, Spain}
\affiliation[3]{Northwestern University, Evanston, IL, USA}
\affiliation[4]{Max-Planck-Institute for Radio Astronomy, Bonn, Germany}
\affiliation[5]{University of California Los Angeles, Los Angeles,  CA, USA}
\affiliation[6]{University of Oxford, UK}
\affiliation[7]{University of Hertfordshire, Centre for Astrophysics Research, College Lane, Hatfield AL10 9AB, UK}
\affiliation[8]{INAF - Osservatorio Astrofisico di Catania, Catania, Italy}
\affiliation[9]{Astrophysics Research Institute, Liverpool John Moores University, Liverpool, L3 5RF, UK}
\affiliation[10]{Cosmic Origins Of Life (COOL) Research DAO, \href{https://coolresearch.io}{https://coolresearch.io}}
\affiliation[11]{University of Oxford, Oxford, UK}
\affiliation[12]{Rhodes University, Makhanda, South Africa}
\affiliation[13]{South African Radio Astronomy Observatory, South Africa}
\affiliation[14]{Institute for Research in Fundamental Sciences (IPM), School of Astronomy, Tehran, Iran}
\affiliation[15]{Department of Astronomy \& Astrophysics, Tata Institute of Fundamental Research,
Homi Bhabha Road, Mumbai 400005, India}
\affiliation[16]{INAF-IAPS, Via Fosso del Cavaliere, 100, 00133 Rome, Italy}
\affiliation[17]{Masaryk University, Department of Theoretical Physics and Astrophysics, Kotl{\'a}\u{r}sk{\'a} 267/2, 611 37 Brno, Czech Republic}
\affiliation[18]{Charles University, Faculty of Mathematics and Physics, Astronomical Institute, V Hole\v{s}ovi\v{c}kách 2, CZ-18000 Prague,
Czech Republic}

\abstract{The Galactic Centre is the nearest nucleus of a galaxy and the most extreme environment that we can observe down to physical scales of a few hundred astronomical units. There is no other region in the Milky Way that can match its unique characteristics, such as its stellar density, turbulence and temperature of the interstellar medium, strong large scale magnetic field, concentration of stellar remnants, or mean star formation rate. The Galactic Centre is a unique target to understand the physics of galactic nuclei and study a large number of rare objects, such as extremely massive stars and stellar remnants, at a well-defined distance. The Galactic Centre has been and is being studied intensively with the most advanced facilities. In this chapter, we advocate for a large-area, multi-wavelength continuum survey with the Square Kilometre Array of an area of about $2.0^{\circ}\times0.4^{\circ}$ ($\sim$ 290\,pc$\times$60\,pc), centred on the massive black hole Sagittarius\,A* and for repeated deep observations of the nuclear star cluster over a decade, which will allow the community to address multiple science problems with a single data set.}

\begin{document}

% Bibliography and bibfile
\def\aj{AJ}%
          % Astronomical Journal
\def\actaa{Acta Astron.}%
          % Acta Astronomica
\def\araa{ARA\&A}%
          % Annual Review of Astronomy and Astrophysics
\def\apj{ApJ}%
          % Astrophysical Journal
\def\apjl{ApJ}%
          % Astrophysical Journal, Letters
\def\apjs{ApJS}%
          % Astrophysical Journal, Supplement
\def\ao{Appl.~Opt.}%
          % Applied Optics
\def\apss{Ap\&SS}%
          % Astrophysics and Space Science
\def\aap{A\&A}%
          % Astronomy and Astrophysics
\def\aapr{A\&A~Rev.}%
          % Astronomy and Astrophysics Reviews
\def\aaps{A\&AS}%
          % Astronomy and Astrophysics, Supplement
\def\azh{AZh}%
          % Astronomicheskii Zhurnal
\def\baas{BAAS}%
          % Bulletin of the AAS
\def\bac{Bull. astr. Inst. Czechosl.}%
          % Bulletin of the Astronomical Institutes of Czechoslovakia 
\def\caa{Chinese Astron. Astrophys.}%
          % Chinese Astronomy and Astrophysics
\def\cjaa{Chinese J. Astron. Astrophys.}%
          % Chinese Journal of Astronomy and Astrophysics
\def\icarus{Icarus}%
          % Icarus
\def\jcap{J. Cosmology Astropart. Phys.}%
          % Journal of Cosmology and Astroparticle Physics
\def\jrasc{JRASC}%
          % Journal of the RAS of Canada
\def\mnras{MNRAS}%
          % Monthly Notices of the RAS
\def\memras{MmRAS}%
          % Memoirs of the RAS
\def\na{New A}%
          % New Astronomy
\def\nar{New A Rev.}%
          % New Astronomy Review
\def\pasa{PASA}%
          % Publications of the Astron. Soc. of Australia
\def\pra{Phys.~Rev.~A}%
          % Physical Review A: General Physics
\def\prb{Phys.~Rev.~B}%
          % Physical Review B: Solid State
\def\prc{Phys.~Rev.~C}%
          % Physical Review C
\def\prd{Phys.~Rev.~D}%
          % Physical Review D
\def\pre{Phys.~Rev.~E}%
          % Physical Review E
\def\prl{Phys.~Rev.~Lett.}%
          % Physical Review Letters
\def\pasp{PASP}%
          % Publications of the ASP
\def\pasj{PASJ}%
          % Publications of the ASJ
\def\qjras{QJRAS}%
          % Quarterly Journal of the RAS
\def\rmxaa{Rev. Mexicana Astron. Astrofis.}%
          % Revista Mexicana de Astronomia y Astrofisica
\def\skytel{S\&T}%
          % Sky and Telescope
\def\solphys{Sol.~Phys.}%
          % Solar Physics
\def\sovast{Soviet~Ast.}%
          % Soviet Astronomy
\def\ssr{Space~Sci.~Rev.}%
          % Space Science Reviews
\def\zap{ZAp}%
          % Zeitschrift fuer Astrophysik
\def\nat{Nature}%
          % Nature
\def\iaucirc{IAU~Circ.}%
          % IAU Cirulars
\def\aplett{Astrophys.~Lett.}%
          % Astrophysics Letters
\def\apspr{Astrophys.~Space~Phys.~Res.}%
          % Astrophysics Space Physics Research
\def\bain{Bull.~Astron.~Inst.~Netherlands}%
          % Bulletin Astronomical Institute of the Netherlands
\def\fcp{Fund.~Cosmic~Phys.}%
          % Fundamental Cosmic Physics
\def\gca{Geochim.~Cosmochim.~Acta}%
          % Geochimica Cosmochimica Acta
\def\grl{Geophys.~Res.~Lett.}%
          % Geophysics Research Letters
\def\jcp{J.~Chem.~Phys.}%
          % Journal of Chemical Physics
\def\jgr{J.~Geophys.~Res.}%
          % Journal of Geophysics Research
\def\jqsrt{J.~Quant.~Spec.~Radiat.~Transf.}%
          % Journal of Quantitiative Spectroscopy and Radiative Trasfer
\def\memsai{Mem.~Soc.~Astron.~Italiana}%
          % Mem. Societa Astronomica Italiana
\def\nphysa{Nucl.~Phys.~A}%
          % Nuclear Physics A
\def\physrep{Phys.~Rep.}%
          % Physics Reports
\def\physscr{Phys.~Scr}%
          % Physica Scripta
\def\planss{Planet.~Space~Sci.}%
          % Planetary Space Science
\def\procspie{Proc.~SPIE}%
          % Proceedings of the SPIE
\let\astap=\aap
\let\apjlett=\apjl
\let\apjsupp=\apjs
\let\applopt=\ao

%Define some commands
%--------------------------------------
\newcommand{\msol}{M$_{\odot}$}
\newcommand{\Msol}{M$_{\odot}$}
\newcommand{\Lsol}{L$_{\odot}$}
\newcommand{\kms}{km~s$\rm ^{-1}$}
\newcommand{\Htwo}{H$\rm {_2}$}
\newcommand{\SFR}{$\rm M_{\odot} yr^{-1}$}
\newcommand{\sgra}{Sgr\,A*}
\newcommand*\degr{\ensuremath{^\circ}}
\newcommand*\arcmin{\ensuremath{^\prime}}
\newcommand*\arcsec{\ensuremath{^{\prime\prime}}}
%--------------------------------------https://www.overleaf.com/project/649d8400c9034aec7909d203

\maketitle

\section{Galactic Centre Science with SKA1-Mid}

The SKA1-Mid array will provide us with exquisite data on the nearest galaxy nucleus and probe the physics of the Milky Way's most extreme environment in unprecedented detail and sensitivity. It will enable us to address a large number of key science cases, such as:

\begin{enumerate}
        \item What is the structure and  formation history of the Galactic Centre (GC) as revealed by its stellar remnants?
        \item How many pulsars are there in the GC and where are they located? Is there a population of millisecond pulsars (MSPs) at the GC that can serve to test General Relativity? Can the presence of MSPs explain the GC  $\gamma-$ray excess?
        \item What are the properties of the winds of massive stars? Is the current day Initial Mass Function (IMF) at the GC different than in the Galactic disk?
        \item What is the present-day star-formation rate at the GC?
         \item How complex can chemistry become in the GC? Can the "building blocks" of life form in GC molecular clouds? 
        \item What is the origin of the magnetized radio filaments in the GC?
        \item What is the activity cycle of accreting Neutron Stars (NSs) and {\bf stellar mass} Black Holes (BHs) in binaries? Can we improve on our understanding of the radio-near infrared emission correlation in Low Mass X-ray Binaries (LMXBs)? 
        \item Can we find rare objects, such as intermediate mass black holes (IMBHs)?
    \end{enumerate}

In addition, auxiliary science cases can be addressed, such as the properties of interstellar extinction towards the GC. The GC is a prime target for all major telescopes with rich synergies for multi-wavelength studies.  At the frequencies that we consider here, SKA1-Mid will provide us with sub-arcsecond angular resolution combined with high sensitivities (see Tab.\,\ref{tab:res-sens}). The southern location of SKA1-Mid implies long observability time windows for the GC  with little elongation of the interferometric beam. It  will therefore enable an efficient deep survey of a large field as well as repeated snapshots of selected pointings for time-domain science and ultra deep studies. 

\begin{center}
\begin{table}[htb]
\centering
\begin{tabular}{lccc}
\hline
Band  & Central frequency  &  Beam  &  Continuum sensitivity\\
    &    [GHz]             &        &       [$\mu$Jy]  \\
\hline\hline
2 &  1.355 & $0.613"\times0.600"$  &  5.26 \\
5a & 6.55  & $0.127"\times0.124"$  &  0.71 \\
5b & 11.85  & $0.070"\times0.069"$  &  0.81 \\
\hline
\end{tabular}
\caption{Beam size and sensitivity of the SKA1-Mid in bands 2, 5a, and 5b. Estimated based on the \href{https://sensitivity-calculator.skao.int/mid}{SKAO Sensitivity Calculator}, assuming subarray configuration AA4 (only 15\,m antennae), default bandwidths and central frequencies, 1\,h of on-source integration time, and Briggs weighting (0). \label{tab:res-sens}}
\end{table}
\end{center}

\section{The Galactic Centre Environment}

Before discussing the science questions more in detail, we will give a brief summary of our current knowledge about the centre of the Milky Way.

The Galactic Centre (GC), at $8.25\pm0.01$\,kpc from Earth
\citep{Gravity-Collaboration:2020oy}, is the only  nucleus of a galaxy
that allows us to study the properties and dynamics of stars and
gas on milli-parsec scales
\citep[e.g.][]{Genzel:2010fk,Schodel:2014bn}. Arguably, no other region on the
sky offers a comparable density of sources that can be studied
indivividually across the electromagnetic regime: A pointing of the SKA1-Mid
towards Sagittarius\,A* (Sgr\,A*), the four million solar mass black
hole at the GC, will contain about $10^7$ stars, a few $10^{5}$ NSs and a few $10^{4}$ stellar mass BHs  within its primary beam \citep[assuming stellar densities and a star
formation history as in][]{Schodel:2018db,Schodel:2020qc}. 

\begin{figure}[!tb]
\center
\includegraphics[width=0.9\textwidth]{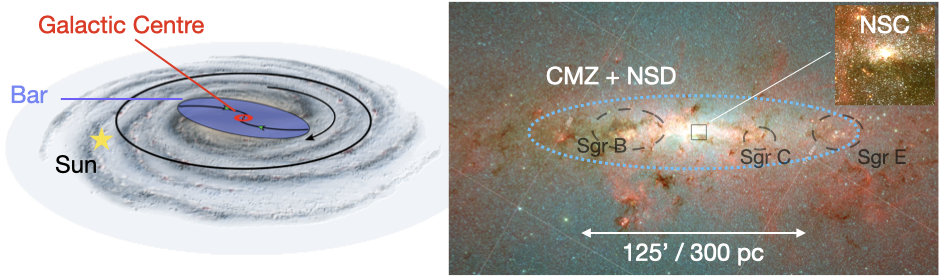}
\caption{\label{Fig:Galaxy} Left: Schematic view of our Galaxy (thanks to M.~Sormani for the sketch). The red circle outlines the location and size
  of the CMZ and NSD. Right: Spitzer $3.6 + 4.5 + 8.0\,\mu$m image of the GC. The NSC surrounds Sgr\,A* and lies embedded in the NSD. The majority of the molecular gas in the CMZ is concentrated within the blue, dotted ellipse, which contains several prominent infrared  dark clouds. In addition to the NSC we have labelled three further regions of special interest with dashed, grey lines: the Sgr\,B molecular cloud complex and star forming region \citep[including the  so-called dust ridge to its west, see][]{Henshaw:2022nm}; and the Sgr\,C and Sgr\,E H\,II regions.
}
\end{figure}

The inward transport of gas through the Milky Way's bar gives rise to
the Central Molecular Zone (CMZ). The conditions in this region
\citep[and in similar ones in the nuclei of nearby
galaxies such as NGC 253, see][]{Sakamoto:2011lj,Martin:2021ot} are so
extreme (in terms of density, magnetic fields, turbulence,
temperature) that it can be considered  our nearest analogue of high-redshift
star-forming regions \citep{Kruijssen:2013tq}.

The GC can serve as a proxy to understand the conditions in CMZs in
general \citep[e.g.][]{Kruijssen:2014ok,Henshaw:2022nm} and it is the
best laboratory where to test whether key prebiotic precursors of the
"building blocks" of life can form in interstellar space
\citep{Jimenez-Serra:2020lo,Jimenez-Serra:2022sw}. It may therefore
 also be an essential piece to understand our cosmic origins.

Figure\,\ref{Fig:Galaxy} shows a sketch of the Milky Way and an overview
of the main components of its central region. (1) The CMZ contains $3-10\%$ of the Milky Way's molecular gas
\citep[][]{Henshaw:2022nm} inside a surface area that is approximately
1000 times smaller than that of the Galactic Disc; (2) the so-called nuclear
stellar disc (NSD) contains about one billion solar masses, with
exponential radial and vertical scale lengths of about 90\,pc
($\sim$38\,arcmin) and 30\,pc ($\sim$13\,arcmin)
\citep{Launhardt:2002nx,Schonrich:2015uq,Schultheis:2019lw,Nogueras-Lara:2020pp,Sormani:2020ve};
(3) the nuclear star cluster (NSC) has an effective radius of $\sim$4\,pc (roughly two arc minutes on the sky) and
a total mass of about $2.5\times10^{7}$\,M$_{\odot}$
\citep{Schodel:2014fk,Feldmeier-Krause:2017rt}. Both the NSC and the NSD have supersolar mean
metallicities and the great majority of their stars appear to be older than 8\,Gyr 
\citep[][]{Nogueras-Lara:2020pp,Schodel:2020qc,Schultheis:2021du,Sanders:2022qa}. 

Averaged over the past 100\,Myr, the CMZ is our Galaxy's most prolific
star forming region \citep{Henshaw:2022nm}, with a mean specific star
formation rate of about
$1\times10^{-6}\,$M$_{\odot}$\,yr$^{-1}$\,pc$^{-2}$ (assuming a star
formation rate of $0.1$\,M$_{\odot}$\,yr$^{-1}$ and a CMZ radius of
300 pc). For comparison, assuming a Milky Way disc radius of 10\,kpc,
the present-day specific star formation rate of the entire Milky Way
is about $1\times10^{-8}$\,M$_{\odot}$\,yr$^{-1}$\,pc$^{-2}$, which is
100 times lower. The extreme conditions in the CMZ  may favour the formation of
massive stars \citep{Morris:1993ve}, and infrared and radio observational studies indeed point to a top heavy initial mass function \citep{Bartko:2010fk,Husmann:2012vn,Lu:2013fk,Hosek:2019vn,Gallego-Calvente:2022al}. 
The Arches and Quintuplet young massive clusters provide a unique opportunity to study the properties and evolution of massive stars \citep[e.g.][]{Clark:2023ua,Cano-Gonzalez:2024wv,Cano-Gonzalez:2025ay}. The presence of massive, young stars in the inner parsec of the NSC provides evidence for in-situ star formation, possibly in a previously existing accretion
disc, a few million years ago.

%\begin{figure}[!hb]
%\center
%\includegraphics[width=\textwidth]{GC_radio.png}
%\caption{\label{Fig:gc_radio} Left: MeerKAT $1.28$\,GHz image of the
%  GC \citep{Heywood:2022rd}.  Galactic
%  north is up, Galactic east is to the left.  Right: Zoom into the
%  central area, that is indicated by the yellow rectangle in the left panel.
%}
%\end{figure}

The dusty, molecular clouds along the line-of-sight to the GC cause
strong attenuation of electromagnetic radiation in the optical to
ultraviolet regime
($\sim$$10^{-12}$ at visible wavelengths). Observations are therefore
limited to the radio/mm, infrared, and
X/$\gamma$-ray regimes. Extreme source crowding requires $\sim$$0.1"$ 
angular resolution to separate the stars and reliably identify
counterparts of sources detected at different wavelengths via
positional cross-matching.

\section{Science cases}

\subsection{Stellar remnants}

It  appears that both the NSC and the NSD formed  early in our
  galaxy's history, at least $8$\,Gyr ago
  \citep{Nogueras-Lara:2020pp,Schodel:2020qc,Sanders:2022qa}. Their
  high masses and old ages imply the existence of large numbers of
  stellar remnants.  We can expect a large population of binaries that contain a stellar remnant to form in the dense GC environment, because the relevant tidal capture mechanisms depend on the square of the stellar density \citep{Voss:2007kb}. In many of those objects the remnant will accrete material from its stellar companion and undergo outbursts. For simplicity, here we will refer to all of these objects as X-ray binaries (XBs). The so-called low mass X-ray binaries (LMXBs), typically contain companion stars of sub-solar masses.

X-ray observations and theoretical predictions indicate the presence of 300 to 600 BH low mass XBs (LMXBs) in the central parsec of the GC \citep[see][]{Hailey:2018eq,Generozov:2018cs,Panamarev:2019gf,Mori:2021jm}. LMXB candidates have also been detected at greater distances, out to  several tens of pc from Sgr\,A*, but at significantly lower densities. However, only two  high mass X-ray binary candidates have been found in the GC \citep{Mandel:2025vn}. 

\begin{figure}[!tb]
\center
\includegraphics[width=0.9\textwidth]{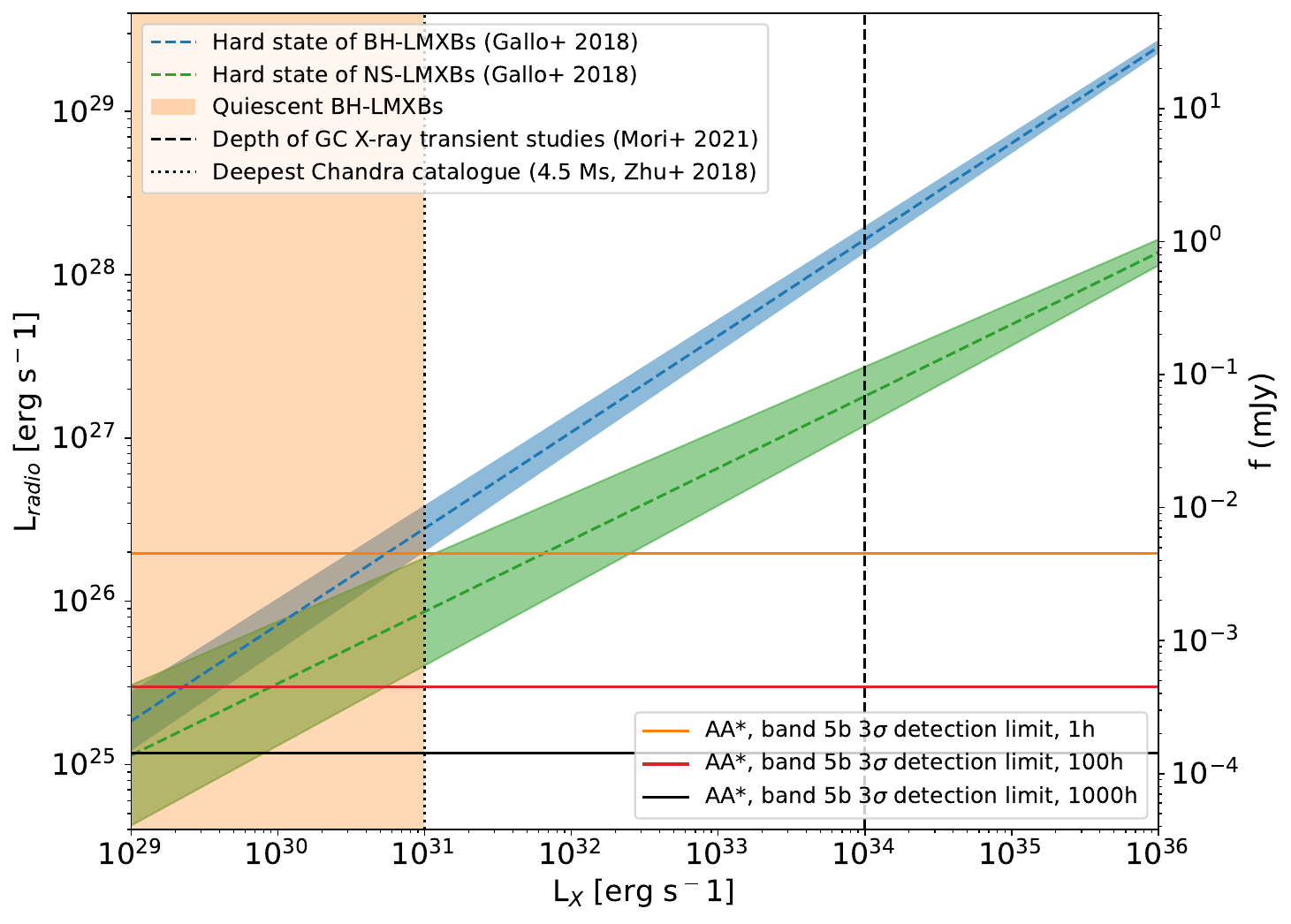}
\caption{ Radio-X-ray correlation for NS and BH LMXBs in the hard state located at the distance of the GC. The blue line/shaded region shows the radio-X-ray correlation for BH LMXBs and the green line/shaded region shows the one for NS LMXBs according to \citet{Gallo:2018ap}.}
\label{Fig:LMXBs}
\end{figure}

Distinguishing between NS and BH LMXBs is of critical importance when studying their numbers, distribution, and properties. With their relatively low masses, NS LMXBs, for example, are less subject to mass segregation and may display a significantly shallower density cusp around Sgr\,A*  than BH LMXBs \citep[see, e.g.,][]{Freitag:2006fk}. With observations at a single wavelength only, the sole criterion for classifying LMXBs in NS or BH categories is the consideration that NS LMXBs are thought to have considerably shorter recurrence times between outbursts ($\lesssim10$\,yr) than BH LMXBs ($>50$\,yr). This is difficult to achieve in the X-ray regime:  Even though routine monitoring of the GC has been carried out with Swift-XRT for over a decade \citep[e.g.][]{Mori:2021jm}, the latter can only reach luminosities down to  $L_{X}\sim10^{34}$\,erg\,s$^{-1}$. The deepest published Chandra catalogue reaches down to $\sim$$10^{31}$\,erg\,s$^{-1}$, close to the expected flux of quiescent LMXBs, but cannot probe the variability of the faint sources (see Fig.\,\ref{Fig:LMXBs}). The angular resolution of X-ray observations is at best $0.5"$\,FWHM, which makes any identification of a stellar counterpart impossible in the crowded GC.

Fortunately, LMXBs can also be observed in the radio regime, where these objects can be detected while being in significantly fainter states and where large arrays can afford us an angular resolution $\lesssim$$0.1"$. At low accretion rates,  in the so called hard state, that is dominated by jet emission, there exists a tight correlation between their radio and X-ray emission \citep[see, e.g.,][]{Fender:2004xv,Corbel:2013ie,Gallo:2018ap}. Since this state extends into very low luminosities and possibly quasi-quiescence, LMXBs spend most of their time in the hard state, where they may be conveniently  detected by radio observations. Indeed, \citet{Zhao:2022wb} claim the detection of about 40 potential LMXBs in the central parsec of the GC with JVLA observations with a resolution of $0.05"$ at $33.0$ and $44.6$~GHz.

\begin{center}
\begin{table}
\centering
\begin{tabular}{llll}
\hline
        &  WDs  &  NSs  & BHs\\
\hline\hline
NSD &  $7.2\times10^{7}$ & $3.5\times10^{6}$  &  $1.8\times10^{6}$ \\
NSC &  $2.6\times10^{6}$ & $1.3\times10^{5}$  &  $6.3\times10^{4}$ \\
\hline
\end{tabular}
\caption{Estimated number of stellar remnants in NSC
 and NSD. \label{tab:remnants}}
\end{table}
\end{center}

We can obtain  estimates of the number of stellar remnants in the GC by assuming the star formation
histories of \citet{Nogueras-Lara:2020pp} for the NSD and of
\citet{Schodel:2020qc} for the NSC. We  approximate the respective
star formation histories with four single-age populations: (1) NSD:
90\% 8\,Gyr, 5\% 1\,Gyr, 4\% 250\,Myr, and 1\%\,40\,Myr, where the
percentages refer to the originally formed stellar mass at the given
age (not to the stellar mass still existing, which is lower). For the total NSD mass we assume $7\times10^{8} \,M_{\odot}$
\citep{Sormani:2020ve}.  (2) NSC: 80\% 10\,Gyr, 15\% 3\,Gyr, 4\%
200\,Myr, and 1\%\,20\,Myr. For its total mass we assume
$2.5\times10^{7} \,M_{\odot}$ \citep{Schodel:2014fk}. Variations in
the exact ages will not affect in any significant way the number of NSs or stellar BHs, and
only to a small degree the number of white dwarfs (WDs). We used the
Stellar Population Interface for Stellar Evolution and Atmospheres
(SPISEA) Python package \citep{Hosek:2020mi} to simulate the stellar
populations and obtain the number of stellar remnants. We assumed
twice-solar metallicity \citep[for metallicities in the GC, see, for
example][]{Rich:2017rm,Nandakumar:2018zr,Nogueras-Lara:2020pp,
  Schodel:2020qc,Schultheis:2021du}. We used the {\it IFMR\_Raithel18}
class to assign initial-to-final mass ratios (IFMRs). Since \citet{Pfuhl:2011uq} find no evidence for a significantly altered IMF in the past, we assumed a power-law IMF constant with time (exponent $-2.3$ between 0.5 and 120 \Msol, $-1.3$ for smaller masses). There is evidence for a top-heavy IMF in the star formation event that created on the order of 200 massive stars within 0.5\,pc of \sgra\ about 4 million years ago \citep[see e.g.][]{Bartko:2010fk,Lu:2013fk}.  Our estimate of the number of stellar remnants may therefore be conservative and their number may be larger if a top heavy IMF has prevailed over the entire life of the GC.

Table~\ref{tab:remnants} lists the estimated numbers of stellar remnants in the NSC and NSD. Different assumptions, such as the
number of star formation epochs, the uncertainties in the originally
formed stellar mass in each epoch, the metallicity, or IFMRs, may
result in changes by up to a few tens of percent, but will not impact the
overall magnitude of the numbers. Assuming a homogeneous distribution
of the remnants throughout the NSC (for simplicity) and an approximate half-mass
radius of 4\,pc (1\,pc corresponds to about
$25''$ at the distance of the GC), a pencil-beam towards the NSC
with a radius of $1.0'$ may contain of the order of 15000 NSs, 7000 BHs, and
$3\times10^{5}$ WDs. These are conservative estimates. Stellar BHs, in particular, are expected to be significantly  concentrated towards Sgr\,A*, forming the so called dark cusp 
\citep[see][]{Morris:1993ve,Baumgardt:2018ad,Hailey:2018eq,Zhao:2022wb,Haas:2025vz}. 

As Fig.~\ref{Fig:LMXBs} shows, with just 1\,h of integration time the SKA1-Mid can detect LMXBs as faint as the faintest sources in the deep Chandra catalogue \citep{Zhu:2018el} and several orders of magnitude fainter than what Swift can detect in its monitoring of transients. Due to its southern location  the SKA1-Mid can monitor the GC conveniently with 1\,h snapshots during long observability windows and with an almost round interferometric beam. The latter is a significant advantage over the JVLA, which requires higher frequencies to reach the necessary angular resolution (where LMXBs will be fainter and observations are deteriorated by stronger atmospheric effects), suffers from beam elongation and more difficult scheduling. By stacking 1000\,h of  observations, we will be able to probe deeply into the quiescent regime of LMXBs. The monitoring should extend over ten years, which covers the recurrence time of NS LMXBs and this will allow us to distinguish them from BH LMXBs, with their much longer recurrence times, and pulsars, which are expected to show little variability.

The corresponding data set will uncover the dark cusp around Sgr\,A* and measure its properties. It will thus allow us to test theoretical stellar dynamics and obtain a constraint on the expected frequency of so-called extreme mass ratio inspiral events (EMRIs), where stellar mass BHs merge with a massive black hole via the emission of gravitational waves \citep[e.g.][]{Amaro-Seoane:2012ty}.  Before the merger, the stellar-mass compact remnants are also expected to interact with the hot flow surrounding Sgr A*, potentially leading to quasiperiodic perturbations of the accretion/outflow rates that can be manifested electromagnetically \citep{Sukova:2021za}.  

As suggested by \citet{Zhao:2022wb}, some of the hyper-compact radio sources in their sample can be powered by accretion of the dense ambient gas onto the stellar remnants. Proper motion measurements will enable us to identify the sources that are co-moving with the ambient gas which are the strongest candidates for such an accretion. Knowing their number will offer us an independent way to estimate the total number of stellar remnants in the dark cusp. The proper motions of the gas can be measured with relative ease within the central parsecs of the GC, as demonstrated, e.g., by \citet{Zhao:2009la} or \citet{Bhat:2022pu}. 

The observed number density of LMXBs will allow us to calibrate their theoretically expected formation rates with observations. Furthermore the number of NS and BH LMXBS will provide us with insight whether the initial mass function at the GC is skewed significantly towards massive stars. By observing large numbers of LMXBs at a well known distance, we will be able to test and calibrate the radio-X-ray correlation of these sources. This is in particular interesting for NS LMXBs, where the latter is less certain than for BH LMXBs. Finally, we will be able to obtain significant constraints on recurrence times, be able to identify pulsars (due to their low variability), and understand whether the so-called Very Faint X-ray transients are BH or NS LMXBs and what their properties are \citep[see, e.g., discussion in][]{Mori:2021jm}.

With even as many as 1000 LMXBs  in the central parsec of the GC  crowding will not pose a problem for the observations. If the LMXBS followed a power law density $\rho\propto r^{-2.75}$, i.e.\ the steepest expected value for the case of strong mass segregation \citep{Alexander:2009gd}, and if we assumed that all of them were visible at the same time, then their surface density would reach about 12\,arcsec$^{-2}$ within a projected radius of  0.01\,pc or $0.25"$ around Sgr\,A*. This could be resolved by the SKA1-Mid in Band 5a. This is an extremely conservative estimate, because the number of LMXBs is expected to be a factor of a few lower \citep[e.g.][]{Generozov:2018cs} and with a significantly flatter distribution \citep[e.g.][]{Baumgardt:2018ad}. Also, not all LMXBs will be active at the same time. Close to Sgr\,A* the brightness of the latter will probably impose limits caused by the limited dynamic range of the observations.

As concerns confusion with other radius sources in the central parsec, those can be radio stars, which are massive stars that can be easily identified, and that have a flat to inverted radio spectrum, or ionised gas, which will also be typically resolved at the high angular resolution of the SKA1-Mid and also has a flat radio spectrum.

When it comes to dynamic-range limited imaging, the interferometric array's theoretical thermal noise is often less important than these limitations. This type of imaging is limited by systematic errors such as calibration errors, deconvolution artifacts, and non-coplanar baselines for large fields of view, among others. The SKA-Mid will have a large number of antennas with impressive uv-coverage, also at the highest frequencies. Techniques such as self-calibration, using multiple compact sources within the Field of View, multi-scale imaging, and advanced weighting can overcome these limitations and enable a high dynamic range of $10^{3} - 10^{4}$ or greater in the study of the Galactic Centre with SKA1-Mid. 

\subsection{Pulsars}

 Pulsars can serve to perform fundamental tests of physics \citep{Kramer:2016uv}.  The study of radio pulsars at the GC is one of the central SKA science cases, because we expect a large number of them in this region and thus a high probability of finding double-pulsars or a pulsar in orbit around a stellar BH. A pulsar may even be present in a tight  orbit around Sgr\,A*. Both scenarios  would provide an exquisite probe of the physics of gravity \citep{Kramer:2004is,Eatough:2015fm}. Even though the probability of observing a normal pulsar (NP) on a short period ($\sim$1\,yr) orbit around Sgr\,A* that is beamed towards Earth may be rather low  \citep{Schodel:2020qc,Zajacek:2018vd}, the intense star-formation activity in the GC in the past $\sim$10\,Myr suggests the presence of (up to a few) $100$ pulsars  throughout the central parsec \citep{Eatough:2015fm,Schodel:2020qc}. About 20\% of them may be beamed towards Earth.

Apart from NPs, which are young NSs with spin periods of one to a few $0.1-1$\,s,
  there exist the so-called millisecond pulsars ({\it MSPs}), old NSs spun up to reach again
  fast rotational periods by accretion from a stellar
  companion. Given the old age of the stellar population in the
  GC as well as the large stellar density, such pairs of
  NSs and donor stars may  form dynamically in close
  encounters. MSPs are of great interest, because they can serve to obtain extremely  accurate dynamics, and therefore insights into the structure and  dynamics of the GC. If they are found near Sgr\,A* or orbiting a stellar BH, they can be exquisite probes of GR/theories of gravity. This is actually true for any pulsar in a sufficiently compact orbit, which is hence a strong motivation to search for pulsars in the GC \citep{Liu:2012zx,Psaltis:2016cu}.
  
Surprisingly, to this day only seven pulsars have been detected inside the CMZ. Six of them lie at projected distances $R>10'$ ($R>24$\,pc) from Sgr\,A* \citep{Deneva:2009fz,Schnitzeler:2016za,Wongphechauxsorn:2024dg}. The seventh one, a magnetar, is at $R$$\sim$$2.5"$ ($R$$\sim$0.1\,pc). 
  
This so-called {\it missing pulsar problem} is a
 key puzzle of GC astrophysics \citep[e.g.][]{Torne:2021hm}. A possible explanation may be the challenge to observe pulsars in this environment: Typically, pulsars and
magnetars are identified by their pulses, but scatter
broadening makes this task difficult at the GC
 \citep[e.g.\ ][]{Eatough:2015fm,Torne:2021hm}. There may also exist
  deeper astrophysical reasons that could explain the dearth of
  pulsars. \citet{Dexter:2014sf} and \citet{Zhao:2022wb}, for example, discuss the intriguing hypothesis that most
NPs may be born as magnetars in the GC. Since the spin-down time
of magnetars is much shorter than the one of NPs, this
hypothesis could explain the missing pulsars problem at the
GC. 

The Galactic Centre $\gamma$-ray excess, discovered by Fermi/LAT more than a decade ago \citep{Hooper:2011aq}, may be caused by the annihilation of dark matter particles at the GC. An alternative explanation for this excess of GeV radiation is the presence of a substantial population of MSPs at the GC. With the SKA we will be able to test this hypothesis with sufficiently deep observations and thus be able to settle a long-standing question.

\citet{Zhao:2020bd} estimate 5.5\,GHz mean flux densities of $50\,\mu$Jy for NPs and $5\,\mu$Jy for MSPs at the distance of the GC. They analysed multi-epoch JVLA  data and, based on the observed luminosity function and non-variability on a timescale of 6 years, classify about  20 of their compact sources as potential NPs. In an analysis of $33.0$ and $44.6$\,GHz JVLA A configuration images of the central $0.8$\,pc$\times0.8$\,pc ($20"\times20"$),   \citet{Zhao:2022wb} find a few dozen compact sources with inverted spectra (i.e., rising towards higher frequencies). They speculate, based on their study of the spectrum of the magnetar SGR\,J1745--2900, that some  of them may be magnetars.

 SKA1-Mid will be instrumental in finding  pulsars at the GC.  It is expected to reach a continuum rms noise of $1.1$, $0.7$, and $0.8\,\mu$Jy\,beam$^{-1}$  at $1.31$, $6.55$, and
$11.85$\,GHz  in one-hour-long integrations (using the SKAO sensitivity calculator and assuming Briggs weighting, see Tab.\,\ref{tab:res-sens}).  Assuming $S_{\nu}\propto\nu^{\alpha}$ and a spectral index of $\alpha=-1.6$ \citep{Jankowski:2018oy}, NPs at the GC will have mean flux
densities of about $510\mu$Jy at $1.4$\,GHz, $40\,\mu$Jy at
$6.7$\,GHz, and $12\mu$Jy at $12.5$\,GHz.  The
corresponding values for MSPs will be $51\mu$Jy at $1.4$\,GHz,
$4\,\mu$Jy at $6.7$\,GHz, and $1.2\mu$Jy at $12.5$\,GHz. 

In 10\,h of observation, an rms noise of $\sim$$0.2\,\mu$Jy\,beam$^{-1}$ can be reached at $6.7$\,GHz, which implies the possibility to detect MSPs at the $7\,\sigma$ level with SKA1-Mid.  Follow-up pulsar-timing, multi-wavelength  and polarisation observations will be essential for confirming the nature of the sources and potentially identifying magnetars via their spectral properties.  Finally, in 100\,h, e.g.\ from stacking ten epochs of 10\,h observations, MSPs can be detected in all three bands with a significance of $>10\,\sigma$, assuming that interstellar scattering does not prevent their detection. Assuming scattering as observed for the Galactic Centre magnetar \citep[see e.g.][]{Spitler:2014qf}, observing frequencies above 30 GHz may be needed. Proper motion measurements can  provide us with upper limits on  how deep pulsars are located inside the potential of
Sgr\,A*. 

Thus, the discovery of a population of pulsars in the Galactic Centre with the SKA1-Mid would represent far more than a neutron star census: such systems would serve as probes for strong field tests of gravity, enable the measurements of the mass distribution and gravitational potential of the inner Galaxy, possibly clarify the origin of the $\gamma$-ray excess, as well as provide insight into the Galactic Centre environment and its stellar evolution history through their characteristic ages and line-of-sight propagation effects.

\subsection{Massive stars}
The GC hosts numerous massive stars, concentrated in the central parsec and in clusters such as the Arches and Quintuplet, but also distributed throughout the NSD \citep[see][]{Clark:2021fj}. Since they all lie at a well-known distance, the GC provides a unique laboratory to study the evolution of massive-stars, in particular as concerns their mass loss. Once off the main sequence, massive stars develop strong winds that produce thermal radio emission, sometimes with non-thermal contributions from colliding winds in binaries. Such radio stars can be easily identified through their precise astrometric coincidence with infrared-bright stars.

\begin{figure}[htb]
  \centering
\includegraphics[width=.8\textwidth]{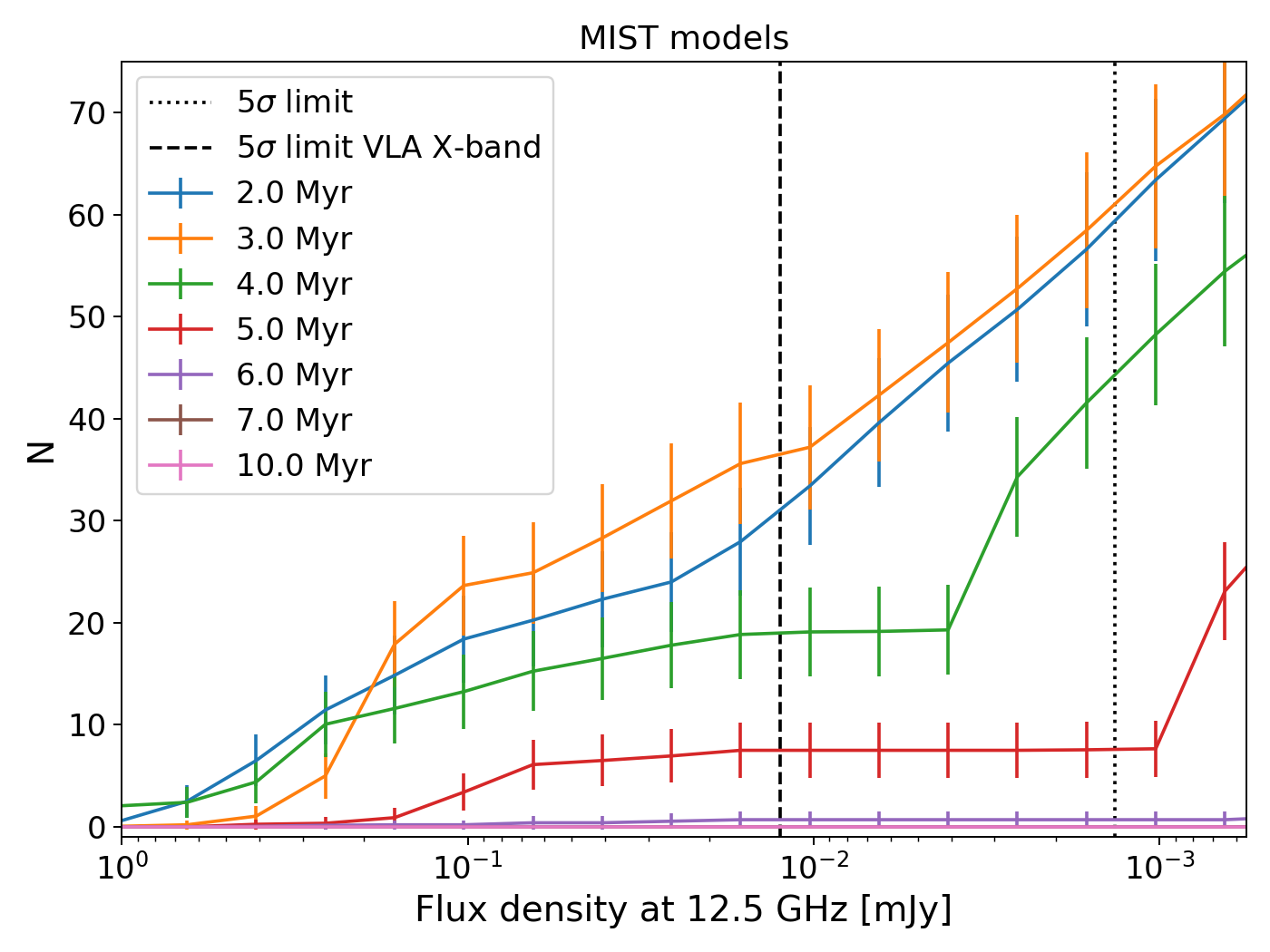}
  \caption{Cumulative radio luminosity functions (RLFs) for
    a star-forming event of $2\times10^{4}$\,M$_{\odot}$ at the distance of
    the GC with different assumed ages and solar metallicity. The prediction are based on MIST
    isochrones, using SPISEA to simulate clusters, assuming an initial mass function with slope $-1.8$ \citep{Hosek:2019vn}. The mass-loss rates of the stars that are provided by
    the MIST models were converted to radio luminosities
    \citep[e.g.][]{Leitherer:1997yf}. The dashed and dotted vertical lines represent the $5\sigma$ detection limit of the VLA in X-band from \citet{Cano-Gonzalez:2024wv}, and SKA1-MID respectively, assuming 10\,h of observing time. The RLFs have different
    characteristic shapes at each age, which can allow us to constrain the age of a star forming event via comparison of the observed flux densities with those derived from synthetic clusters. Further details on the simulations can be found in Cano-Gonz{\'a}lez et al.\ (subm.\ to A\&A).
    }
  \label{fig:RLF}  
\end{figure}

Obtaining a complete census of young massive stars in the GC is challenging due to severe extinction and crowding \citep{Schodel:2014bn}. Even though massive post-MS stars have been detected across the GC via narrow-band infrared, radio, and X-ray observations \citep[e.g.][]{Mauerhan:2010kb,Dong:2011ff},  existing surveys remain incomplete and have great difficulties in detecting O-type main-sequence stars. An alternative is to identify massive stars through their strong ionised winds or circumstellar nebulae, depending on evolutionary state \citep[e.g.\ Pistol star near the Quintuplet cluster, e.g.][]{Lang:2005zr}.  

The most sensitive radio studies so far targeted the Arches and Quintuplet clusters \citep{Gallego-Calvente:2021yl,Gallego-Calvente:2022al,Cano-Gonzalez:2024wv}, demonstrating that radio observations can constrain wind mass-loss rates, multiplicity, and cluster ages. The radio luminosity function provides a novel probe of the initial mass function, while long-term monitoring of colliding-wind binaries can reveal orbital properties. Recent work indicates that additional young massive clusters remain to be identified in the GC \citep{Nogueras-Lara:2022jz,Martinez-Arranz:2024gm} and can provide new targets for radio studies.  

With its $\sim$5$\times$ higher angular resolution, SKA1-Mid will resolve crowded regions (Arches, Quintuplet, central parsec) and deliver precise astrometry for kinematic measurements. At 12.5\,GHz, a 10\,h observation can reach an rms noise of $0.4\,\mu$Jy\,beam$^{-1}$, enabling detections of winds from all post-MS stars and even O main sequence stars down to $\sim$30\,M$_\odot$. This will allow systematic studies of wind mass loss from hundreds of stars at a well determined distance. Simultaneous observations of several hydrogen recombination lines will constrain the internal structure and dynamics of stellar winds via 3D radiative-transfer modelling \citep[using the MOdel for REcombination LInes or MORELI code][]{Baez-Rubio:2014te,Martinez-Henares:2023cy},  probing their launching mechanisms and energetics in statistically relevant samples. 

Figure~\ref{fig:RLF} illustrates the diagnostic power of the radio continuum luminosity function. For example, the Sgr\,B1 region may contain $\sim10^{5}$\,M$_\odot$ of stars formed $\sim$10\,Myr ago \citep{Nogueras-Lara:2022jz}. Since the radio luminosity function evolves rapidly with age, it can place strong constraints on the ages of young clusters and associations in the GC.

The progress offered by the detection and measurement of the winds from massive stars by SKA1-Mid will be twofold. First, the strongest winds are emitted by massive stars after they evolve off the main sequence. This part of stellar evolution is still poorly constrained due to the relative rareness of such stars, that are often located a large and not well constrained distances. With SKA1-Mid we will be able to study up to several hundred of such stars at a well constrained distance. Second, more young massive clusters will be detected with upcoming proper motion studies \citep[see][]{Martinez-Arranz:2024gm,Martinez-Arranz:2024nr,Martinez-Arranz:2025dt}. The radio luminosity functions of these clusters will serve to constrain their properties, in particular their ages \citep[see Fig.\,\ref{fig:RLF}, Cano-González et al., subm.\ to A\&A, and][]{Gallego-Calvente:2022al}.

\subsection{Present day star formation} 

The GC appears to be a significant outlier in the empirically derived relations between gas density and star formation activity \citep[see review by][]{Henshaw:2022nm}. One of the main difficulties in constraining the present-day star formation rate is that the extreme opacity of the molecular clouds at the GC may prevent us from
  detecting a significant fraction of the deeply-embedded young stellar objects.  SKA1-Mid will cover a series of rotational lines of cyanopolyynes such as HC$_3$N, HC$_5$N, HC$_7$N, and HC$_9$N \cite[e.g.][]{Bianchi:2023fh} which are widely detected in the Galactic Centre \citep[e.g.][]{Mills:2018kt}. These molecules are excellent probes of the deeply-embedded population of massive young stellar objects at the hot core stage\footnote{Hot cores are compact ($\leq$0.1 pc), dense (H$_2$ volume gas densities $\geq$10$^8$ cm$^{-3}$) and hot (T$\geq$150-300 K) condensations that represent the cradles of massive stars \citep{Garay:1999pr}.}, where they are  enhanced  by the thermal evaporation of ices from grains \citep{Viti:2004to,Martin-Pintado:2005zt}. HC$_5$N is particularly interesting for SKA1-Mid because it presents within bands 5a and 5b several rotational lines in the ground vibrational level v=0 and in the vibrationally-excited states v$_{10}$ and v$_{11}$. The high-angular resolution of SKA1-Mid will filter out any low-excitation HC$_5$N extended emission, thus pin-pointing the location of the high-excitation HC$_5$N emission arising from hot cores. Therefore, molecular line observations of HC$_5$N (and, possibly, of even larger cyanopolyynes) with SKA1-Mid  will provide information about the youngest population of massive stars in the making in the GC, complementary to continuum imaging which will identify later evolutionary stages such as HC/UCHII regions.

 Last but not least, SKA1-Mid will also be able to probe maser emission from high-mass star formation (including methanol masers in bands 5a and 5b). This would not only be useful to identify regions of high-mass star formation, but could additionally mark targets for SKA-VLBI astrometry and parallax measurements, along the lines of what the BeSSeL survey has provided for the northern sky (e.g., \citealp{Reid:2014jk}). The line emission of HC$_3$N and HC$_5$N has recently been imaged across the CMZ within the ALMA Large Program ACES (ALMA CMZ Exploration Survey). This survey, however, has been acquired with an average angular resolution of 1.5$"$, insufficient to resolve the hot core population whose expected hot core sizes are $500-1000$\,AU ($\sim 0.06^{\prime\prime}$ to $0.13^{\prime\prime}$ at the distance of the GC).   
  
SKA1-Mid will cover all rotational lines from J=5 to J=2 in the ground vibrational level of HC$_5$N, plus its vibrationally excited lines. HC$_5$N is measured to be factors of 2-3 less abundant than HC$_3$N in GC sources \citep[see][]{Zeng:2018hb}. For the hot cores in the Sgr B2 cloud, the HC$_3$N column densities are of $\sim$10$^{17}$ cm$^{-2}$ within a beam of 1.5$"$ \citep[see Table 8 in][]{de-Vicente:2000sh}. Note that this value is a lower limit since the HC$_3$N column densities measured for the hot cores distributed between Sgr B2(N) and Sgr B2(M) \citep[with much weaker emission than the Sgr B2(N) and (M) hot cores themselves; see Figures 1 and 3b in][]{de-Vicente:2000sh} are $\geq$2$\times$10$^{17}$ cm$^{-2}$. When considering that hot cores have actual sizes of $\sim$0.1$"$, these column densities translate into $\geq$10$^{19}$ cm$^{-2}$. Therefore, we can safely assume a HC$_5$N column density of $\sim$3$\times$10$^{18}$ cm$^{-2}$ for the hot cores in the GC. Using this column density, a linewidth for the HC$_5$N emission of 8 km s$^{-1}$, and a gas temperature of 300 K \citep[typical of hot cores in the GC;][]{de-Vicente:2000sh}, we expect peak fluxes for the HC$_5$N rotational lines of $\sim$125-300 $\mu$Jy/beam. These lines will be detected with S/N$>$3-7 in integrated intensity, unveiling the distributed population of GC hot cores.

\subsection{Star formation at the dust-lane CMZ intersection}

\noindent Gas inflow along bar dust lanes is a universal feature of barred galaxies, where gas streams intersect circumnuclear rings and trigger shocks and collisions, but with widely varying star formation efficiencies. In M83, ALMA shows enhanced shock and dense-gas tracers at both bar-ring junctions, yet only one forms massive stars \citep{2019ApJ...884..100H}.  A similar asymmetry exists in the Milky Way: the far-side junction hosts the Sgr~E complex with $\gtrsim$60 compact H\,II regions \citep{2020ApJ...901...51A}, while the near-side junction near G1.3 is quiescent \citep{2022A&A...668A.183B}. The Sgr~E regions are unusual in showing similar radio luminosities and sizes but weak mid-IR PDRs, possibly due to envelope stripping by orbital motion. Current surveys (GBT, MeerKAT, etc.) are biased toward luminous O-star HII regions.  

% With $\mu$Jy sensitivity and sub-arcsecond resolution, SKA1-Mid will extend this census to fainter H\,II regions and use recombination lines to confirm membership. Continuum imaging and spectral-index mapping will uncover diffuse ionised emission and distinguish thermal from non-thermal components, while long ionised tails will constrain stripping or wind-driven processes. 

Sensitive radio continuum emission will enable the identification and characterization of even the youngest phases of H\,II regions, the so-called Hyper-compact and Ultra-compact HII regions even within the GC region. Together with radio recombination lines \citep[RRLs][]{Karska01.2026.SKA} to confirm membership, SKA1-MID continuum imaging and spectral-index mapping will uncover diffuse ionised emission and distinguish thermal from non-thermal components, while long ionised tails will constrain stripping or wind-driven processes. The identification of HII regions will be possible by the proposed continuum survey of the Galactic Plane at 10-15 GHz with SKA1-Mid \citep{Traficante01.2026.SKA}, with a sensitivity of 20 $\mu$Jy at 15 GHz. The GC Survey will provide even more sensitive data and spectral indices over a wider frequency range.

SKA1-Mid will thus provide the first complete inventory of star formation at a bar-CMZ junction, clarify why some intersections (e.g.\ G1.3) remain quiescent, and place the GC in the broader context of star formation at bar--ring junctions.

\subsection{Searching for the ``building blocks'' of life}

The Giant Molecular Clouds in the GC are the most chemically rich repositories of molecular material in our Galaxy \citep[e.g.][]{Requena-Torres:2008rw}. In the past few years, about two dozen new molecular species have been discovered in the dense clouds in the CMZ. Interestingly, some of these species are of prebiotic interest and include precursors of ribonucleotides, nucleobases, sugars, proto-proteins and proto-lipids \citep[e.g.][]{Rivilla:2021qc,Rivilla:2022mw,Rivilla:2023bd,Rodriguez-Almeida:2021ud,Rodriguez-Almeida:2021if,Zeng:2021kc,Sanz-Novo:2023hs,Jimenez-Serra:2020lo,Jimenez-Serra:2022sw}. The CMZ is a suitable environment to detect these molecules because of the large column densities found in its molecular clouds. Therefore, it presents us with a unique opportunity to learn about how complex the chemistry of the ISM can become, and whether the building blocks of life could already form in interstellar space. 

All these molecules have been discovered in emission in the 7\,mm, 3\,mm and 2\,mm bands and they likely only represent the tip-of-the-iceberg. Due to the high sensitivity already reached by  single-dish, broadband spectral surveys we are starting to reach the line confusion limit in the observed spectra for these objects, which prevents the discovery of new prebiotic species of even higher complexity. The only way to circumvent this problem is to observe these molecules at even lower frequencies, which are much cleaner from simpler and smaller molecules. The problem is that the line transitions of large molecules such as prebiotic species become weaker since their partition functions become large (due to the spread over a great number of energy levels). In addition, at low frequencies their Einstein A$_{ul}$ coefficients become orders of magnitude smaller  (typically from 10$^{-6}$ s$^{-1}$ to 10$^{-9}$ s$^{-1}$). This issue is alleviated if we observe these species in absorption against strong background sources instead of in emission \citep{Jimenez-Serra:2022cr}. 
 
By performing spectroscopic measurements toward strong continuum centimeter sources, SKA1-Mid has the potential to discover new prebiotic interstellar species in absorption spectra. As an example, \citet{Jimenez-Serra:2022cr} have shown that a few rotational lines from the C3 sugar glyceraldehyde will be detected with a S/N$\geq$3 in integrated intensity with SKA1-Mid in bands 5a and 5b in 1 hour integration time. Larger sugars such as erythrulose (a C4 ketose sugar), however, will require hundreds of hours of integration time \citep{Jimenez-Serra:2022cr}. Nevertheless, the survey proposed here will represent a key step for the discovery of large prebiotic molecules, because it will reveal the location of the strongest continuum background sources towards which to carry out follow-up dedicated high-sensitivity spectral studies with the SKA1-Mid.   

\subsection{Origins and impacts of the magnetic fields in the GC}%Properties and physics of the large-scale magnetic field}

%\subsubsection{\bf Properties and physics of the large-scale magnetic field}
Among the biggest open questions concerning the GC magnetic field are:
(1) Is the milligauss vertical field homogeneous and pervasive
throughout the CMZ and does this define an overall GC
magnetosphere or are the strongest magnetic features only localised,
transient features? (2) How and where does the vertical field couple
to the horizontal field? Can we find specific points of interaction?
(3) What is the origin of the relativistic electrons that light up the
non-thermal filaments (NTFs)? (4) What is the origin of the poloidal
field? What is the impact of the magnetic fields on the energy balance and structure formation in the GC region?

 Studies of MeerKAT continuum
 data at 1.3\,GHz \citep[e.g.][]{Heywood:2022rd,Yusef-Zadeh:2023zt,2026ApJ...997...31M} have provided an impressive new view of the GC
 magnetic field by tracing thermal and non-thermal filaments with unprecedented sensitivity. Still, the data  are limited to an angular resolution
 $\simeq 5^{"}$, which makes it hard to untangle different features and
 detect faint filaments.

 Full polarization radio continuum observations allow for studying the structure and strength of the magnetic field. The structure of the field, as well as the strength of the regular field along the line of sight, are studied through rotation measure (RM) synthesis. The magnetic field strength in the plane of the sky, which is dominated by turbulent field, is estimated by mapping the pure synchrotron emission \citep{2026ApJ...997...31M}. 
 In bands 2, 5a, and 5b the SKA1-Mid will provide sensitive, high-resolution radio continuum maps,
 enabling to study the magnetic field with angular resolutions almost ten times better than the published MeerKAT observations. This will
 shed light on the role of the magnetic field in the GC on scales down
 to milli-parsecs. So far undiscovered faint filaments will improve
 our knowledge of the magnetic field and of  the origin of the
 relativistic electrons that power the NTFs. One particularly important
 question here is the existence or not of non-thermal horizontal filaments close
 to the Galactic plane. The data will provide us with a better
 understanding of the influence of the magnetic field on the dynamics
 of the ISM, which may even come to dominate some HII
 regions \citep[see][]{Bally:2024iu}. An exciting prospect is that the
 high angular resolution of SKA-MID may allow us to perform proper
 motion measurements of magnetic filaments over timespans of a few to
 ten years. Proper motions could help us clarify the relative locations,
 interactions and kinematics of the vertical field, in particular whether its rotation around the GC resembles the one of  the ISM and stars, which could indicate whether we are
 seeing a global, long lived field configuration or whether the NTFs are localised, short-lived features.

\subsubsection{\bf Origin of the non-thermal filaments}

One of the first hints that the nucleus of our Galaxy harboured energetic activity was the discovery of the archetype magnetised radio 
filaments in the GC 40 years ago \citep{Yusef-Zadeh:1984dz}. Since then, radio observations have revealed a large population of linearly polarised  synchrotron emission tracing nucleus-wide cosmic ray activity throughout the inner few hundred parsecs of the Galaxy \citep[e.g.][]{Heywood:2019fu}. 
%Furthermore, H$_3^+$ absorption-line measurements of this region determined high cosmic-ray ionisation rates indicating that relativistic particles permeate the CMZ at levels a thousand times that in the solar neighbourhood \citep{Oka:2005ef,Indriolo:2012bv,Le-Petit:2016qf}. These particles provide a significant source of pressure in the GC when compared to thermal gas pressure in the ISM of the GC.  
A population of radio filaments has also been detected in radio galaxies in poor clusters.   These extragalactic filaments show remarkably similar underlying physics to those of the GC, in spite of vastly different scale lengths and environments \citep{Yusef-Zadeh:2023zt} and their origin is not understood  \citep{Ramatsoku:2020ei,Rudnick:2022on}.  However, the  similarities provide an opportunity to compare the physical processes in the ISM of the GC to the ones in the  intracluster environment  of poor clusters. 

 There is no obvious source that powers these 
non-thermal filaments.  Numerous models have been proposed to explain their origin, but there is no consensus.  In the cometary scenario, for example, the filaments result from the collection and draping of magnetic field lines by  an energetic cosmic ray source, such as 
a  stellar wind bubble or a pulsar that moves with respect to the medium, thus forming a cometary tail 
 \citep{Yusef-Zadeh:2019pt}. The recent discovery of the first GC MSP \citep{Lower:2024jh}, that is likely associated with a filament, provides support to the hypothesis that pulsars power the NTFs.

A search for compact radio sources associated with the NTFs based on MeerKAT data found an initial sample of 46 sources close to the ends of 
filaments \citep{Yusef-Zadeh:2023zt}. Currently, the nature of these compact sources and their 
physical association with the filaments are not clear.  SKA1-Mid  will be able to test the cometary model. In particular, the spectral index of the compact sources will place constraints on their nature (pulsars or stellar wind sources).  In addition, high spatial resolution will provide details of the possible interaction between
the compact  sources with the filaments.  Lastly, the remarkable sensitivity of SKA1-Mid will be able to uncover fainter filaments than those 
detected by MeerKAT.

\subsubsection{Impact of the magnetic fields and nonthermal pressures}  

The centre of the Milky Way provides a unique laboratory for studying the formation of molecular clouds and stars within a strongly magnetized interstellar medium (ISM). Such studies are particularly crucial for understanding the remarkably low star formation efficiency observed in this region \citep{Longmore:2013yu}. Recent observations highlight the dynamical importance of non-thermal pressure components—introduced by magnetic fields and cosmic-ray electrons—in shaping the ISM \citep{Nasirzadeh:2024mw,Hassani:2022sv}. In particular, magnetic fields can facilitate the onset of galactic outflows \citep{Tabatabaei:2022vk}, support molecular clouds against gravitational collapse, and slow down the formation of massive stars \citep{2018NatAs...2...83T}.

Using MeerKAT observations, \citet{2026ApJ...997...31M} showed that molecular clouds within a projected distance of 7 pc from Sgr A$^{\star}$ are magnetically subcritical, indicating that their evolution is largely regulated by magnetic fields. Future SKA-Mid observations in Bands 1, 2, and 5 will enable a more detailed investigation of the interplay between the magnetized ISM and cloud and star formation in the GC.

\subsection{The base of the Galactic Chimney}

%One of the most intriguing large-scale vertical features of the CMZ is the Galactic chimney \citep{Ponti:2019ul,2019Natur.573..235H,2021A&A...646A..66P}, extending few hundred parsecs above the plane. Its apparent base lies offset from Sgr A*, between the Sgr A and Sgr C regions, and its driving mechanism is not yet clear. If powered by stellar feedback, it could represent the collective output of massive star clusters; if AGN-like in origin, it could be a fossil remnant of enhanced SMBH activity within the past few Myr. SKA1-Mid is ideally suited to address this question by analysing the sources and emission near the chimney’s base. Continuum imaging across bands 2, 5a, and 5b will allow us to disentangle thermal and non-thermal emission, identify compact radio sources (e.g. pulsars, stellar remnants, IMBH candidates), and measure spectral indices. Polarisation studies will probe the connection between local magnetic fields and the vertical structure, while radio recombination line measurements will trace the ionised gas. Together, these data will clarify the nature of both the compact sources and extended emission, and test whether the chimney reflects feedback from distributed star formation, relic SMBH activity, or a combination of both.

One of the most intriguing large-scale vertical structures in the CMZ is the Galactic chimney \citep{Ponti:2019ul,2019Natur.573..235H,2021A&A...646A..66P,Veena:2023ev}, extending a few hundred parsecs above the Galactic plane. Its apparent base is offset from Sgr A*, and its physical origin remains uncertain. Proposed driving mechanisms broadly fall into two categories: starburst (SB)-driven feedback, including clustered supernovae and stellar winds, and AGN-driven fossil activity associated with past outbursts of the central supermassive black hole (SMBH). The synchrotron cooling time inferred for the chimney emission ($\sim$1--2 Myr) implies a recent or sustained injection of relativistic particles, consistent with both scenarios. In SB-driven origin, cosmic rays are expected to be injected in a spatially distributed manner by clustered supernovae, implying that residual tracers of enhanced massive star formation such as pulsars, compact supernova remnants, and evolved or disrupted H II regions should still be detectable near the base of the chimney. Sensitive SKA1-Mid RRL observations can reveal faint or fossil ionised gas structures even if bright classical H II regions have already dispersed, whereas an AGN-driven origin does not require an accompanying stellar or compact-source population. 

Although both scenarios generate non-thermal synchrotron emission, they differ in particle injection and transport. An SB-driven chimney is expected to show a radio spectral index that steepens progressively with distance from the Galactic plane due to cosmic ray transport and energy losses, whereas an SMBH-driven outflow should exhibit steep-spectrum emission already at the base and a more coherent, vertically ordered magnetic field anchored to the nucleus. Multi-band SKA1-Mid continuum and polarisation measurements will enable spatially resolved tests of these predictions (Section 3.7). The chimney base also provides a direct link to the origin of the NTFs (Section 3.8). A census of compact radio sources combined with high-resolution imaging of filamentary structures will test whether the relativistic particle population is predominantly associated with distributed stellar sources (e.g. pulsars or wind bubbles) or instead reflects injection from a central engine, thereby distinguishing SB-driven and SMBH-driven origins and constraining possible hybrid scenarios.

\subsection{Intermediate mass black holes}

Intermediate-mass black holes (IMBHs) have masses in the range between stellar-mass black holes and massive black holes in galactic nuclei, i.e. $m_{\bullet}\sim 10^2-10^5\,M_{\odot}$. They are thus crucial to understand black hole growth and evolution across the cosmic history. So far there is an apparent gap in the mass distribution of black holes in the IMBH mass range due to the difficulty of their detection \citep[see e.g.][for a review]{Greene:2020pl}. This stems from their small region of gravitational influence  and a low luminosity of the accretion flow surrounding them \citep{Labaj:2025fd} in the quiescent state. 

The Galactic center region is expected to be a promising region for hosting IMBHs. Due to their relatively short dynamical friction timescale, $\tau_{\rm df}\propto \sigma_{\star}^3/(m_{\bullet} \rho_{\star})\sim 10^5-10^6$ years, IMBHs should concentrate towards the SMBH. As many as $\gtrsim 50 (m_{\bullet}/150\,M_{\odot})^{1/2}$ IMBHs could accumulate towards Sgr~A* during the cosmic history \citep{Madau:2001jw}. They are generally thought to form primordially \citep[the collapse of population III stars, direct gas cloud collapse,][]{Loeb:1994rv,Eisenstein:1995kt,Madau:2001jw,Begelman:2006th,Latif:2013lm} or in dense star clusters by black hole-black hole mergers or black hole-star collisions  \citep{Miller:2002pb,Portegies-Zwart:2002ow,Fragione:2022bd,Rose:2022wh}.

The SKA can detect IMBHs via the radio emission of the optically thick, self-absorbed synchrotron component of their SED \citep{Hosseini:2024jp,Peissker:2024ly,Labaj:2025fd}. For the case when an IMBH is surrounded by a handful of stars, the IMBH of $\sim 10^4\,M_{\odot}$ would be detectable at the flux density of $\sim 130\,{\rm \mu Jy}$ at 10 GHz \citep{Labaj:2025fd}, which is well within the sensitivity limits for the SKA 5b configuration.

To confirm the mass of detected compact sources, it will be necessary to complement radio observations with mid- and near-infrared data to constrain their SEDs. While for Sgr A*, the spectrum of the SED peaks at $\sim 1$ mm, for IMBHs it is expected in the mid-infrared domain \citep{Labaj:2025fd}, while for isolated stellar black holes, it peaks towards the near-infrared part of the spectrum, assuming an advection-dominated accretion flow.

\section{Properties of compact radio sources at the GC}

\begin{figure}[!htb]
\center
\includegraphics[width=\textwidth]{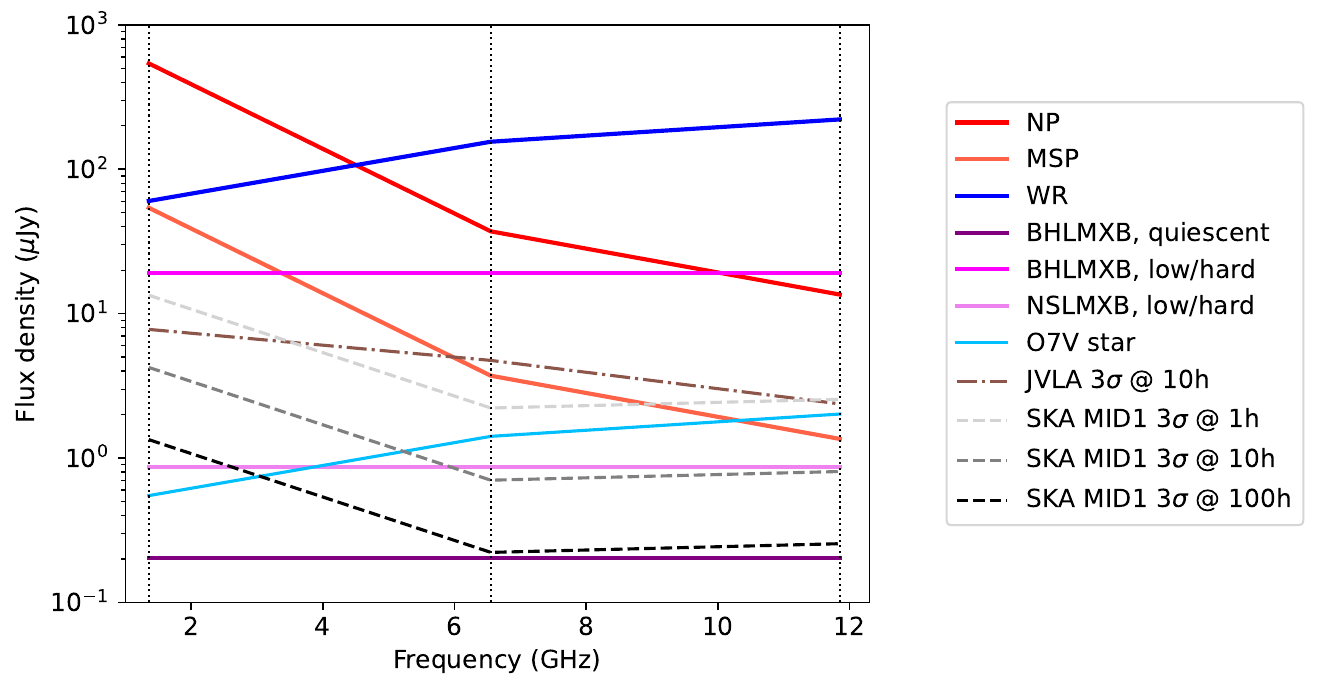}
\caption{\label{fig:sources} Flux densities of different types of
  radio point sources that we can expect to detect at the GC (solid lines). 
  Vertical dotted lines indicate the central frequencies of bands 2
  (1.355\,GHz), 5a (6.55\,GHz), and 5b (11.85\,GHz). The dashed gray to black
  lines indicate the $3\,\sigma$ rms of SKA1-Mid for
  integration times of 1\,h, 10\,h, and 100\,h. The dash-dotted brown line indicates the $3\,\sigma$ rms of the current JVLA. A GC distance of
  $8.25$\,kpc was assumed \citep{Gravity-Collaboration:2020oy}. NP: Normal pulsar,
  assuming a mean flux density of $50\,\mu$Jy at 5\,GHz and a spectral index of
  $-1.7$ \citep{Zhao:2020bd}.  MSP: Millisecond pulsar,
  assuming a mean flux density of $5\,\mu$Jy  at 5\,GHz  and a spectral index of
  $-1.7$ \citep{Zhao:2020bd}. WR: Wolf Rayet star, assuming a flux
  density of of $200\,\mu$Jy  at 10\,GHz  and a spectral index of
  $0.6$ \citep{Gallego-Calvente:2021yl}. O7V star: Thermal wind from an
  O7V star (approximately 30\,M$_{\odot}$) assuming a mass loss rate
  of $1\times10^{-7}$\,M$_{\odot}$yr$^{-1}$, based on a MIST version
  1.2 isochrone with solar metallicity and assuming a thermal spectral
  index of $0.6$. BHLMXB quiescent: Here we have used the observed flux density of the quiescent BHLMXB XTE J1118+480, observed by \citet{Gallo:2014ce}, scaling it to the distance of the GC and assuming a flat spectral
index. BHLMXB, low/hard: BH low mass X-ray binary in low/hard state,
assuming a flat spectral index and a radio/X-ray luminosity of
$10^{28}$\,erg\,s$^{-1}$/$10^{34}$\,erg\,s$^{-1}$. These fluxes correspond to the faintest detections given in 
\citet{Gallo:2018ap}, i.e.~we expect most sources to be brighter. NSLMXB, low/hard: NSLMXB in low/hard state,
assuming that NSLMXBs are about 22 times less radio loud than BHLMXBs \citep{Gallo:2018ap}. As for BHLMXBs, this is a lower limit.
%Many of these sources will be detectable with 1 hour of observing time, with the possibility of revisiting them in different observing runs in a very efficient way. It will also be possible to stack all  images, which will significantly improve the detection thresholds.
}
\end{figure}

Figure\,\ref{fig:sources} provides an overview of the expected flux densities of point-like sources at the GC. The different types of point sources can be distinguished in various ways: Massive stars will have bright near-infrared counterparts. Pulsars will typically have steep spectra \citep[with the possible exception of magnetars, which may have flat to inverted spectra, see][]{Zhao:2022wb}, polarised emission \citep{Sobey:2022tn}, and show little variability. LMXBs will show great variability with most of them manifesting as transient sources. In the low/hard state they typically show flat to inverted spectra from optically thick jet bases, i.e. $\alpha>0$ for $S_{\nu}\propto\nu^{\alpha}$, where $S_{\nu}$ is the radio flux density and $\nu$ the frequency,  but they become optically thin ($\alpha<0$) in the high/soft state \citep{Fender:2004xv}. As concerns multi-wavelength cross-identification, Ultracompact HII and similar nebular sources  have sizes $\lesssim0.1$\,pc, corresponding to $\lesssim2.5"$ at the distance of the GC. Practically all of them will therefore be resolved and/or have an infrared counterpart. Near-infrared imaging with an angular resolution of $0.2"$ is available from the GALACTICNUCLEUS survey \citep{Nogueras-Lara:2019yj}.  Higher angular resolution imaging from JWST NIRCam has already been obtained on some selected fields (such as the central parsecs, Sgr\,C, or the Brick molecular cloud) and may become available for a significant fraction of the GC in the next years \citep{Schoedel:2023rf}. 

\section{Requirements and possible setup for a survey of the GC}

The science cases described above could be addressed with a  survey that covers three aspects: (1)  a wide field that encompasses the entire CMZ; (2) observations at the highest sensitivity  and angular resolution of the environment of Sgr\,A*; (3) a  time domain study of the NSC and a comparison field in the NSD to constrain properties of stellar remnants, such as the recurrence time between LMXB outbursts, and massive stars, such as the variability of their winds, and to measure  proper motions.

\begin{figure}[!t]
  \centering
    \includegraphics[width=\columnwidth,angle=0]{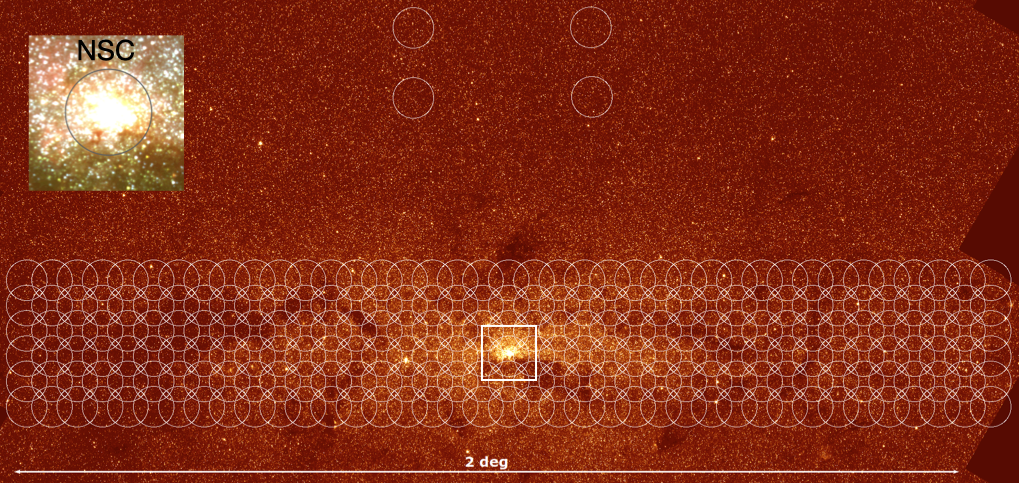}
  \caption{Possible layout of pointings at
  $12.5$\,GHz, overlaid over a Spitzer $3.6\,\mu$m image. Galactic
  north is up, Galactic east is to the left. The NSC is shown in the
  zoomed inlet, with the 11.85\,GHZ primary beam overplotted. The four
  disconnected circles to the Galactic north are comparison
  fields in the inner Galactic bar.}
  \label{fig:pointings125} 
\end{figure}

Multi-wavelength observations will be necessary because spectral
indices provide us with insight into the nature of the observed sources. The SKA GC survey should be carried out at frequencies of $1.36$, $6.55$, and $11.85$\,GHz (bands 2, 5a, and 5b). The angular resolution at these frequencies will be $0.6", 0.13"$, and $0.07"$, respectively. This means that at the highest frequency and with a ten-year time baseline the proper motions of point sources detected at 10\,$\sigma$ can be measured with a precision  of 1\,mas\,yr$^{-1}$. 
%The proper motions will serve to constrain to which stellar structure a source pertains \citep[e.g., see][]{Shahzamanian:2022vz} and how deep it is inside the gravitational potential at the GC.  
In addition, polarisation measurements will be crucial to confirm non-thermal sources such as NSs or BHs. 

We assume that the full width at half maximum (FWHM) of the  SKA1-Mid primary beam is $2.6'$  at $12.5$\,GHz. To ensure a largely homogenous signal to noise ratio across the field we plan to space the pointings in steps of corresponding to the FWHM at each frequency. Assuming 1\,h observations per pointing, we can cover a
$2.0^{\circ} \times 0.4^{\circ}$ field oriented parallel to the
Galactic Plane and centred on Sgr\,A* in 234\,h, as is shown in
Fig.\,\ref{fig:pointings125}. 

The same field can be covered in about 70\,h at $6.7$ GHZ and 5\,h at $1.4$ GHz.  In total, we will require 325\,h for a survey that reaches an rms noise in the continuum images of $2.0$, $1.3$, and $1.2\,\mu$Jy\,beam$^{-1}$ at $1.4$, $6.7$, and $12.5$ GHz, respectively, assuming a fractional bandwidth $\Delta \nu / \nu$ = 0.3. For the line images, the expected rms noise level will be 140, 90 and 85 $\mu$Jy\,beam$^{-1}$, respectively, assuming a fractional bandwidth $\Delta \nu / \nu$ = $10^{-4}$ and a spectral resolution of $322.56$ kHz.  

This shallow survey should be complemented by repeated deep
observations of the NSC.
%, which is the astrophysically most interesting region of the GC. 
We propose to observe the NSC  with 1\,h snapshot imaging in bands 5a and 5b. 
The snapshots should be repeated over at least five to ten years (approximately the recurrence time of NS LMXBs) to allow us to observe variability and proper motions. The final depth of the data (through stacking) should reach $\sim$1000\,h,  sufficient to detect quiescent XBs at $>5\sigma$ in band~5.
%The total time requirement will be 300\,h. 
%of which 100\,h at $12.5\,$, 30\,h at $6.7\,$ and 10\,h at $1.4\,$ GHz).

A comparison field in the NSD should be observed to understand the contamination of the NSC field by sources in the fore-, background, and in the NSD. We consider 100\,h sufficient for this comparison field, because the main purpose will be to understand contamination, which can probably be extrapolated from the luminosity function of the faint sources. We also assume that the density of stellar remnants will be closely related to the stellar density, which is about five times higher in the central parsec than in the NSD \citep{Gallego-Cano:2020jq}. The comparison field can be pointed toward the young Arches or Quintuplet clusters of massive clusters. This will allow for the sampling of the radio emission from its massive stars down to late type O-stars and studying their variability, thus enabling unique science. The comparison field will also allow us to study how the occurrence of LMXBs scales with stellar density. In conclusion, the science cases laid out in this chapter could be addressed by about  1425\,h  of observing time  with SKA1-Mid. 

\section*{Acknowledgements}
AA, RS, MPT, MCG, JM, AG, LVM, SSE, AMA and FNL acknowledge financial support from the Severo Ochoa
 grant CEX2021-001131-S funded by MCIN/AEI/ 10.13039/501100011033. AMA, RS, AG, and FNL acknowledge financial support from grant
 PID2022-136640NB-C21. 
 AA, MPT and JM acknowledge financial support from the Spanish grant PID2023-147883NB-C21, funded by MCIU/AEI/ 10.13039/501100011033, as well as support through ERDF/EU.
 FNL gratefully acknowledges financial support from the Ramón y Cajal programme (RYC2023-044924-I), funded by MCIN/AEI/10.13039/501100011033 and FSE+ and from grant PID2024-162148NA-I00, funded by MCIN/AEI/10.13039/501100011033 and the European Regional Development Fund (ERDF) “A way of making Europe”. MZ acknowledges the support of the Czech Science Foundation Junior Star grant no. GM24-10599M. I.J-.S acknowledges funding from grant PID2022-136814NB-I00 funded by the Spanish Ministry of Science, Innovation and Universities/State Agency of Research MICIU/AEI/ 10.13039/501100011033 and by “ERDF/EU”, and from the ERC grant OPENS (project number 101125858) funded by the European Union. LVM acknowledges financial support from the grants: CEX2021-001131-S funded by MICIU/AEI/ 10.13039/501100011033, PID2021-123930OB-C21 and PID2024-155817OB-I00 funded by MICIU/AEI/ 10.13039/501100011033 and by ERDF/EU, as well as from "The coordination of the participation in SKA-SPAIN", funded by the Ministry of Science, Innovation and Universities (MICIU) and from RED2022-134464-T funded by MICIU/AEI/ 10.13039/501100011033.

\bibliographystyle{abbrvnat-maxbibnames4.bst}
\bibliography{BibGC_SKA} % if your bibtex file is called example.bib

\end{document}